\documentclass[english,british]{IEEEtran}
\usepackage[T1]{fontenc}
\usepackage[latin9]{inputenc}
\usepackage{float}
\usepackage{amsmath}
\usepackage{amssymb}
\usepackage{graphicx}
\usepackage{esint}

\makeatletter

\floatstyle{ruled}
\newfloat{algorithm}{tbp}{loa}
\providecommand{\algorithmname}{Algorithm}
\floatname{algorithm}{\protect\algorithmname}

\makeatother

\usepackage{babel}
\addto\captionsenglish{\renewcommand{\algorithmname}{Algorithm}}

\begin{document}

\title{Bidding policies for market-based HPC workflow scheduling}

\author{Andrew Burkimsher, Leandro Soares Indrusiak\\\{andrew.burkimsher, leandro.indrusiak\}@york.ac.uk\\ Department of Computer Science, University of York, Heslington, York, YO10 5GH, UK}
\maketitle
\begin{abstract}
This paper considers the scheduling of jobs on distributed, heterogeneous
High Performance Computing (HPC) clusters. Market-based approaches
are known to be efficient for allocating limited resources to those
that are most prepared to pay. This context is applicable to an HPC
or cloud computing scenario where the platform is overloaded. 

In this paper, jobs are composed of dependent tasks. Each job has
a non-increasing time-value curve associated with it. Jobs are submitted
to and scheduled by a market-clearing centralised auctioneer. This
paper compares the performance of several policies for generating
task bids. The aim investigated here is to maximise the value for
the platform provider while minimising the number of jobs that do
not complete (or starve).

It is found that the Projected Value Remaining bidding policy gives
the highest level of value under a typical overload situation, and
gives the lowest number of starved tasks across the space of utilisation
examined. It does this by attempting to capture the urgency of tasks
in the queue. At high levels of overload, some alternative algorithms
produce slightly higher value, but at the cost of a hugely higher
number of starved workflows. \end{abstract}


\section{Introduction}

In recent years, computational performance has increasingly been achieved
through increasing parallelism rather than increased single-core performance.
This is mainly due to the exponential increase in power consumption
required to run processors at higher clock speeds \cite{intel04}.
Processors have become multicore, and multicore processors have been
grouped into clusters and networks of clusters, known as grids \cite{kesselman98}.
These High Performance Computing (HPC) systems have grown to be huge
in scale, especially those that support the operations of cloud computing
corporations.

Many kinds of work run on HPC systems are not independent. Instead,
workflows (or \emph{jobs}) are run that are made up of several individual
pieces of software (\emph{tasks}), each of which takes some input
and produces some output. This leads to the notion of \emph{dependencies}
between tasks, which restrict tasks to starting only once all their
dependencies have completed and their input data transmitted.

Large-scale systems such as these can experience periods of high demand.
This can even extend into periods of overload, where work is arriving
faster than it can be executed. HPC providers need to be able to effectively
prioritise work during these periods to ensure the most important
work is run. Cloud providers who sell computing capacity on the open
market will wish to maximise their profits by running the work that
is most valuable \cite{cloudintro11}. There are real costs to HPC
providers for running work, so it may sometimes be more worthwhile
to discard work whose value is too low than to run it.

Work is run on computing clusters because it is valuable to the users
and organisations who submit it. This value can be represented as
a real amount of money that users are willing to pay for their results.
This value is not always fixed. This is because work returned earlier
may allow users or organisations to be more productive, whereas some
work may have no value at all if it is returned too late \cite{aburkimsherEngD14}.

Scheduling large-scale computing systems has traditionally been done
through list scheduling \cite{graham69,Maheswaran99}. This is where
all the work in a queue is ranked by an appropriate metric, and allocated
to hardware in sorted order. Where value is assigned to work, market-based
principles can be applied to scheduling.

The usefulness of work on list scheduling can be employed in a market-based
way. The values calculated for ranking in a list scheduler can instead
be used as bids that each piece of work submits to an auctioneer.
Past research has investigated using value as bids \cite{irwin04}.
However, the bids can be different than just the values of jobs at
a given moment in time. It has been shown \cite{aburkimsherEngD14}
that in the context of list scheduling, ordering tasks by total job
value does not necessarily give the highest overall value under overload.

This paper explores the effectiveness of different bidding policies
in a market-based scheduling scenario. This is to evaluate which provide
the highest value, especially in periods of overload when not all
work is able to run.

\section{Literature Survey}

Scheduling using market-based techniques has been studied over a long
period, with pioneering early work by Sutherland in the 1960s \cite{Sutherland68}.
Since then, a great deal of work has been done on the structure of
markets, investigating which structures best promote economic efficiency.
Auction-based markets have been looked at by Waldspurger et al. \cite{waldspurger02}
whereas market-clearing techniques where resources are allocated to
work until one or other is consumed were investigated by Miller and
Drexler \cite{millerdrexler88}. Lai et al. have examined doing the
allocation of resources in proportion to the value of bids \cite{Lai05admissons}.
Popovici and Wilkes \cite{popoviciwilkes05} investigated market-inspired
admission control. Dube \cite{dube08} gives a good survey of the
state of this research and concludes that various different market
structures all hold promise, showing the benefits of better allocation
and greater decentralisaion (and hence scalability and fault-tolerance)
than traditional schedulers.

A necessary part of a market based system is that jobs have a certain
value to users, and that this value is represented in the market by
their bids \cite{irwin04}. There has been much past work on scheduling
to maximise the value of a workload \cite{Lee99}. The problem of
maximising workload value where job value is a function of time was
extensively researched by Locke \cite{Locke1986} for systems schedulable
by the Earliest Deadline First list scheduling policy, and has subsequently
been applied to more general scheduling problems \cite{aburkimsherEngD14,chen96}.

It seems that there has been little research into using value curves
rather than fixed job values when scheduling in a market-based context.
This is likely because markets need a single price for each bid submitted
by a job, not a curve. Yet in previous work \cite{aburkimsherEngD14},
it has been shown that value curves can be used to create single ranking
values for list scheduling at given times in a simulation. This paper
will extend this approach by using these values as bids in a market-based
framework.

\section{Models}

To evaluate which bidding policies are best, the simulator previously
developed by the authors \cite{aburkimsherEngD14} was extended to
use a market-based scheduling system. This simulator implements several
models representing the applications, the platform, the scheduling
scheme and the means of representing and calculating value. The simulation
takes place in a discrete-time environment, where all events happen
on time ticks $\tau\in\mathbb{N^{\mathrm{0}}}$.

\subsection{Application Model}

The application model represents the work to be run on the cluster.
Tasks are non-pre-emptive, running to completion once they have begun.
This represents the behaviour of the LSF/GridEngine systems commonly
used to manage large HPC systems. This lack of pre-emption is less
of a problem at scale than it might seem, because in large-scale systems,
the turnover of work is sufficiently high. This means that it is very
unlikely that any job will ever have to wait too long for something
to finish so that it can start \cite{burkimsher14}. 

For the purposes of this paper, a single, non-preemptible piece of
work will be known as a \emph{task}, denoted $T^{i}$. Each task will
run for a duration $T_{\mathtt{exec}}^{i}\in\mathbb{N^{\star}}$ on
a number of cores $T_{\mathtt{cores}}^{i}\in\mathbb{N^{\star}}$ concurrently.
A set of tasks is known as a \emph{job}, denoted $J^{k}$. Tasks can
only depend on other tasks within the same job and the successors
of each task will be known as $T_{\mathtt{succ}}^{i}$ . The structure
of the dependencies will be that of a Directed Acyclic Graph (DAG),
which defines a topological ordering over the tasks in a job: a partial
order that they must be run in.

Having a dependency graph means several other useful metrics about
a job can be calculated. The upward rank \cite{Topcuoglu2002} of
tasks, $T_{\mathtt{R}}^{i}$, is defined as the longest route from
any task to its latest-finishing successor: $\forall\,T^{j}\in T_{\mathtt{succ}}^{i}:T_{\mathtt{R}}^{i}=T_{\mathtt{exec}}^{i}+\max(T_{\mathtt{R}}^{j})$.
Sorting tasks by their upward rank will give an ordering that respects
the partial order of the dependencies. The upward rank is also useful
to estimate the finish time of a job if a task is run immediately.
As jobs only realise their value once they have finished, this estimate
is useful for scheduling comparisons at runtime.

The critical path \cite{kelley61} of a job, $J_{\mathtt{CP}}^{k}$,
is equivalent to the largest upward rank of any of its component tasks:
$\forall\,T^{i}\in J_{\mathtt{}}^{k}:J_{\mathtt{CP}}^{k}=\max\left(T_{R}^{i}\right)$
\cite{Topcuoglu2002}. This is the length of time that the job would
take to execute if the number of cores requied was unbounded, which
is equivalent to its minimum execution time.

\subsection{Platform Model}

The basic unit of the platform model is an execution core. Each cluster
is made up of a number of cores. Multicore tasks must execute within
a single cluster, and consume a number of cores for the duration of
their execution. Cores are not shared between executing tasks. Within
a cluster, cores are assumed to have negligible communication delays
between them. For example, this could be due to all cores sharing
a single networked file system. In the HPC context considered, execution
times are measured in hours to days, so the highest-speed communications
within cores on the same multicore processor or within the same server
can reasonably be assumed to be negligible.

Clusters communicate through a central router. Delays are present
for data transfers between clusters, and are adjustable using the
Communication to Computation Ratio (CCR), supplied as an input to
the simulations. For a task executing for $T_{\mathtt{exec}}^{i}$,
the time taken for data transfer would be $T_{\mathtt{exec}}^{i}\times CCR$.

Heterogoneity is present in the model to a limited degree. Tasks and
cluster cores have \emph{Kinds}, which must match for a task to be
able to run on a cluster. A task will always run for the same amount
of time on clusters of the same kind. Each cluster is made up of cores
of only one kind. This means that for jobs where tasks require different
kinds, some network communications are unavoidable, as some tasks
must run on different clusters. Where this is the case, the network
delays are considered in the calculation of the critical path as well
as just the execution times.

\subsection{Scheduling Model}

The scheduling model is market-based and works by using a central
auctioneer running on the central router. All jobs are submitted to
the central router, which maintains the queue of work. Jobs are immediately
decomposed into their component tasks and, once their dependencies
are satisfied, are added to the single global queue. This is different
to previous work by the authors \cite{aburkimsherEngD14}, where work
is load-balanced on submission and then spends time queueing on the
clusters.

Scheduling takes place at each scheduling instant, which can be the
arrival of a job or the completion of a task on any of the clusters.
At each instant, the tasks and the clusters submit bids to the auctioneer.
The tasks bid according to a bidding policy, and these bidding policies
will be described below. The clusters bid according to the number
of cores they have available. The auctioneer then assigns the highest-bidding
task to the cluster with the most cores free until there are no tasks
left or there are no clusters with sufficient cores free to run the
highest-bidding task. This is an example of a market-clearing architecture.

When a task is assigned to a cluster, it is run immediately, because
there are guaranteed to be sufficient cores free on a cluster to run
due to the bidding mechanism.

Backfilling is not used where tasks other than the highest bidder
could fit onto a cluster even where the highest bidder cannot. This
is for several reasons. Most importantly, when running large-scale
clusters, there is a high turnover of work. This means that the delay
until the next running task finishes is likely to be small \cite{burkimsher14}.
This means there is less likely to be much of a penalty without doing
backfilling. Secondly, where execution times are only estimates, backfilling
cannot be perfectly precise. This may mean that some previously lower-bidding
tasks are still running when the previously high-bidding task would
have started. This can delay the execution of the high-bidding task,
penalising the overall value achievable.

\subsection{Value Model}

Users submit work to HPC systems because running the work on their
own computers is likely to take a prohibitively long time. Furthermore,
users tend to not really care how busy the systems are overall, but
instead care most about the responsiveness of their jobs. Therefore,
a model of value curves is required that captures these perceptions
about responsiveness. It is assumed that users will define these curves
and submit them along with their jobs. This is because the decision
and allocation of appropriate time and value is context-dependent
and is fundamentally a stakeholder issue.

The authors previously concluded \cite{burkimsher12} that the most
appropriate measure of responsiveness for jobs with dependencies was
Topcuoglu's Schedule Length Ratio or \emph{SLR} \cite{Topcuoglu2002}.
This is the ratio of the job's actual response time to the length
of its critical path: $SLR=\left(J_{\mathtt{finish}}^{k}-J_{\mathtt{arrive}}^{k}\right)\div J_{\mathtt{CP}}^{k}$.
A particularly useful feature of the SLR is that job SLR can be estimated
(or projected, hence P-SLR) for tasks in advance if the submission
time of the job, the current time and the upward rank of the task
is known: $T^{i}\in J^{k}:\;PSLR\left(T^{i}\right)=\left(\left(T_{R}^{i}+\tau_{\mathtt{current}}\right)-J_{\mathtt{arrive}}^{k}\right)\div J_{\mathtt{CP}}^{k}$.

Each job submitted will also be submitted with a value curve, $J_{C^{i}}^{K}$.
The curve $C^{i}$ is an array of coordinates of the points that define
the curve. The time-axis of the value curve is defined using the SLR
of the job. This means that the initial and final deadlines along
with the time coordinates of points on the curve are defined in terms
of SLR. This is so that the value curve can easily be scaled to jobs
of different sizes, total values or lengths of critical path. This
is also designed so that the future value of a task can be estimated
using its P-SLR. The value curve is undefined before an SLR of 1,
as no job can finish before the length of its critical path. 

\begin{figure}
\noindent \begin{centering}
\includegraphics[bb=0bp 0bp 523bp 282bp,clip,width=1\columnwidth]{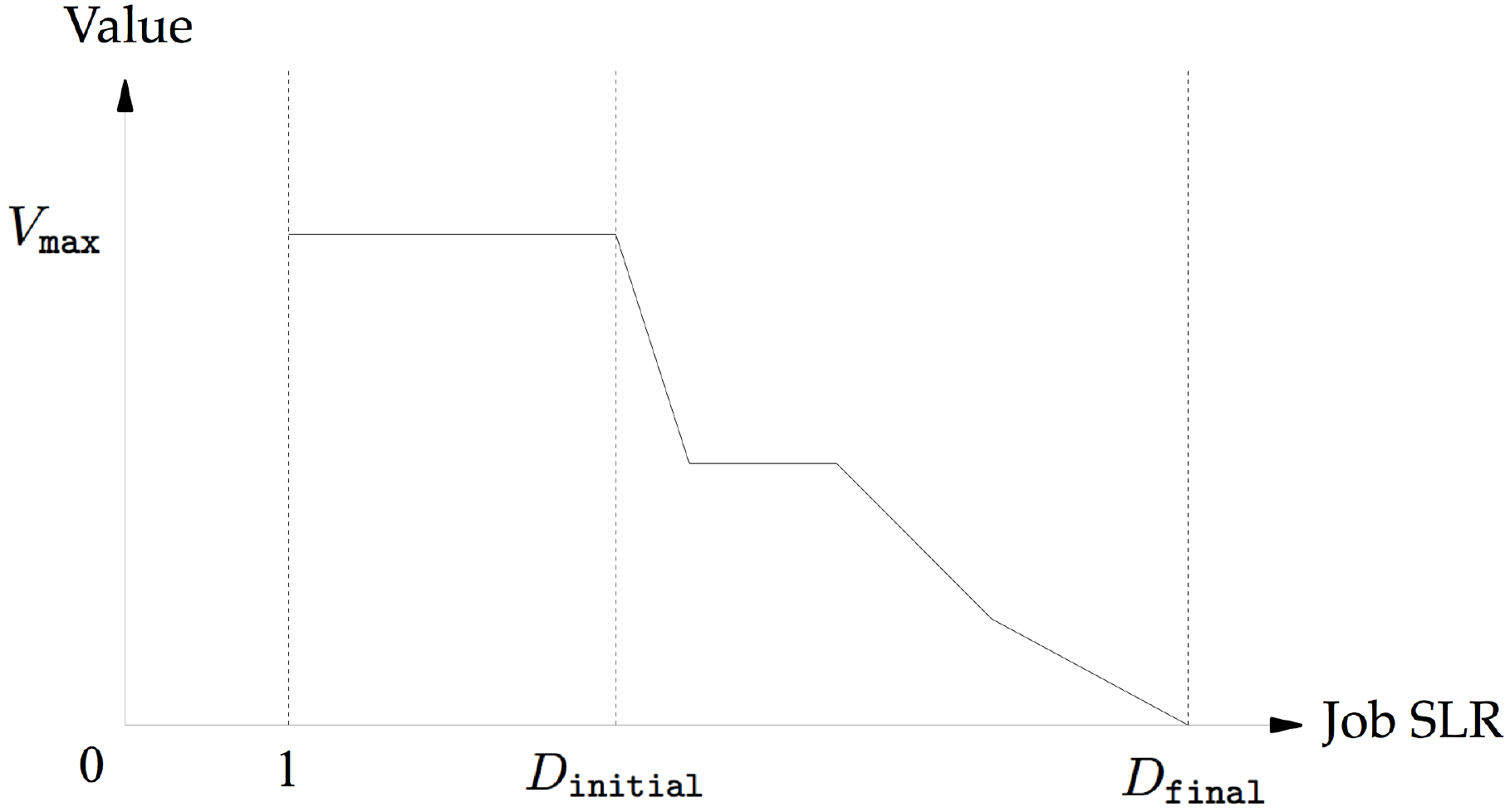}
\par\end{centering}

\protect\caption{Value Curve Template}

\label{value-curve-template-pic}
\end{figure}

Figure \ref{value-curve-template-pic} shows the template used to
define value curves. Every job is assigned a maximum value $V_{\mathtt{max}}$
that it can return to the user. A value curve is defined as a piecewise
function with three subdomains, punctuated by an initial ($D_{\mathtt{initial}}$)
and a final ($D_{\mathtt{final}}$) deadline, as defined in Algorithm
\ref{value-algorithm}. Before the initial deadline, the value returned
is always the maximum $V_{\mathtt{max}}$. Between the initial and
final deadlines, the value is calculated using linear interpolation
between a sequence of points which reach 0 at $D_{\mathtt{final}}$.
Once the final deadline has been reached, the value is 0. The algorithm
for calculating this factor is given in Algorithm \ref{value_factor_algorithm}.

\begin{algorithm}
$\mathtt{Value}\left(\textrm{job}\;J^{k},\mathtt{SLR}\right)=$

\[
\begin{cases}
J_{\mathtt{VMax}}^{k} & \mathtt{if\;SLR}\leq D_{\mathtt{initial}}\\
0 & \mathtt{if\;SLR}\geq D_{\mathtt{final}}\\
J_{\mathtt{VMax}}^{k}\times\mathtt{factor}\left(J_{C^{i}}^{k},\:\mathtt{SLR}\right) & \mathtt{if}\;D_{\mathtt{initial}}<\mathtt{SLR}<D_{\mathtt{final}}
\end{cases}
\]

\protect\caption{Value Calculation}

\label{value-algorithm}
\end{algorithm}

\begin{algorithm}
$\mathtt{factor}\left(\textrm{curve}\;C^{i},\mathtt{SLR}\right):$

\quad{}$\mathtt{t\_sort=}\left[\mathtt{p}\left[0\right]\mathtt{\:for\:p\:in\:}C^{i}\right]$

\quad{}$\mathtt{v\_sort=}\left[\mathtt{p}\left[1\right]\mathtt{\:for\:p\:in\:}C^{i}\right]$

\quad{}$\mathtt{low\_index=}\left|\left[\mathtt{t\:for\:t\:in\:t\_sort\:if\:t}\leq\mathtt{SLR}\right]\right|$

\quad{}$\mathtt{t\_low=t\_sort}\left[\mathtt{low\_index}\right]$

\quad{}$\mathtt{t\_high=t\_sort}\left[\mathtt{low\_index}+1\right]$

\quad{}$\mathtt{v\_low=v\_sort}\left[\mathtt{low\_index}\right]$

\quad{}$\mathtt{v\_high=v\_sort}\left[\mathtt{low\_index}+1\right]$

\quad{}$F=\mathtt{v\_low}+\left(\frac{\mathtt{SLR-t\_low}}{\mathtt{t\_high-t\_low}}\times\left(\mathtt{v\_high-v\_low}\right)\right)$

\quad{}$\mathtt{return}\:F$

\protect\caption{Value Factor Calculation\label{value_factor_algorithm}}
\end{algorithm}

\section{Bidding Policies}

In a market, independent agents need to make bids for the services
they require. In this context of market-based scheduling, tasks will
each bid for computational resource. In this market, it is necessary
to have policies that calculate the bids for the tasks. These bids
can be based a number of underlying attributes of the task.

Baseline policies are explicitly intended to be simplistic or naive,
to show what is achievable with little processing work. Two baseline
policies are considered, \emph{Random} and \emph{First In First Out}
(FIFO). Random places a random bid for each task in the queue at each
round of bidding. FIFO places bids that are proportional to the arrival
time of each task's parent job $T^{i}\in J^{k},\;FIFO=J_{\mathtt{arrive}}^{k}$,
with the smallest numerical bid being the highest priority.

Several policies are considered that do not make use of the value
curves. These instead use the upward ranks of tasks $T_{R}^{i}$ within
the dependency graph of their parent job. \emph{Shortest Remaining
Time First} (SRTF) and \emph{Longest Remaining Time First} (LRTF)
\cite{zhaosakellariou06,Topcuoglu2002} bid for tasks using their
upward rank, with the smallest and largest bids being given priority,
respectively.

The \emph{Projected Schedule Length Ratio} (P-SLR) policy bids the
P-SLR value, with the highest value being highest priority \cite{burkimsher12}.
This is so that the tasks that are most late relative to their execution
time are given the highest priority, which is intended to ensure that
all jobs will have a waiting time proportional to the length of their
critical path, a desirable attribute for fairness and responsiveness
\cite{saule10}. The P-SLR as defined above is starvation-free as
long as overloads are transient (overall load is below 100\%). However,
to ensure P-SLR is starvation-free under extremely high load, a further
term is added to the equation. The equation as used for P-SLR bidding
in the simulation is shown in Algorithm \ref{Proj-slr-algo}. One
discrete time unit is also added to the P-SLR calculation in order
to preferentially prioritise smaller tasks that arrive at the same
instant as larger ones.

\begin{algorithm}
$\mathtt{projected\_slr}(\textrm{task}\;T^{i},\;\textrm{job\ }\;J^{k},\mathtt{curr\_time},\;\textrm{queue}\;Q)=$
\[
\frac{\left(T_{R}^{i}+\mathtt{curr\_time}+1\right)-J_{\mathtt{arrive}}^{k}}{J_{\mathtt{CP}}^{k}}+\left\lfloor \frac{\mathtt{curr\_time}-J_{\mathtt{arrive}}^{k}}{\forall J^{n}\in Q:\max\left(J_{\mathtt{CP}}^{n}\right)}\right\rfloor ^{2}
\]

\protect\caption{\selectlanguage{english}%
Projected SLR algorithm\selectlanguage{british}%
}
\label{Proj-slr-algo}
\end{algorithm}

Several policies based on value are considered. \emph{Projected Value}
(PV) uses the estimated P-SLR to determine the parent job's value
if it were to finish after the task's upward rank. PV then uses the
highest bids as the most preferable. This is similar to previous market-based
policies where work has fixed values rather than value curves $T^{i}\in J^{k}:\;PV\left(T^{i}\right)=\mathtt{Value}\left(J^{k},\;PSLR\left(T^{i}\right)\right)$.

The issue with PV is that tasks that promise a large amount a value
may also take a large amount of resource. Instead, the \emph{Projected
Value Density} (PVD) policy \cite{Locke1986} looks at how profitable
running each task may be, by comparing the value gained with the resource
required to achieve this value. The resource required, $T_{\mathtt{succ\_sum}}^{i}$,
is the sum of the execution times, $T_{\mathtt{exec}}^{i}$, multiplied
by the cores they require, $T_{\mathtt{cores}}^{i}$, of all the tasks
that depend on the considered task $\forall\,T^{j}\in T_{\mathtt{succ}}^{i}:T_{\mathtt{succ\_sum}}^{i}=\left(T_{\mathtt{exec}}^{i}\times T_{\mathtt{cores}}^{i}\right)+\sum T_{\mathtt{succ\_sum}}^{j}$.
Having defined the resource required, we can define the value density
as the value divided by the resource required $T^{i}\in J^{k}:\;PVD\left(T^{i}\right)=\mathtt{Value}\left(J^{k},\;PSLR\left(T^{i}\right)\right)\div T_{\mathtt{succ\_sum}}^{i}$
. Higher density is prioritised as it will be most profitable.

Aldarmi and Burns \cite{aldarmi1999} suggested that squaring the
value density gives a better separation between tasks that will be
profitable and those that should be starved to avoid wasting resources.
This is termed the \emph{Projected Value Density SQuared} (PVDSQ)
policy, $PVDSQ\left(T^{i}\right)=\left(PVD\left(T^{i}\right)\right)^{2}$.

\begin{figure}
\noindent \begin{centering}
\includegraphics[bb=0bp 0bp 525bp 220bp,width=1\columnwidth]{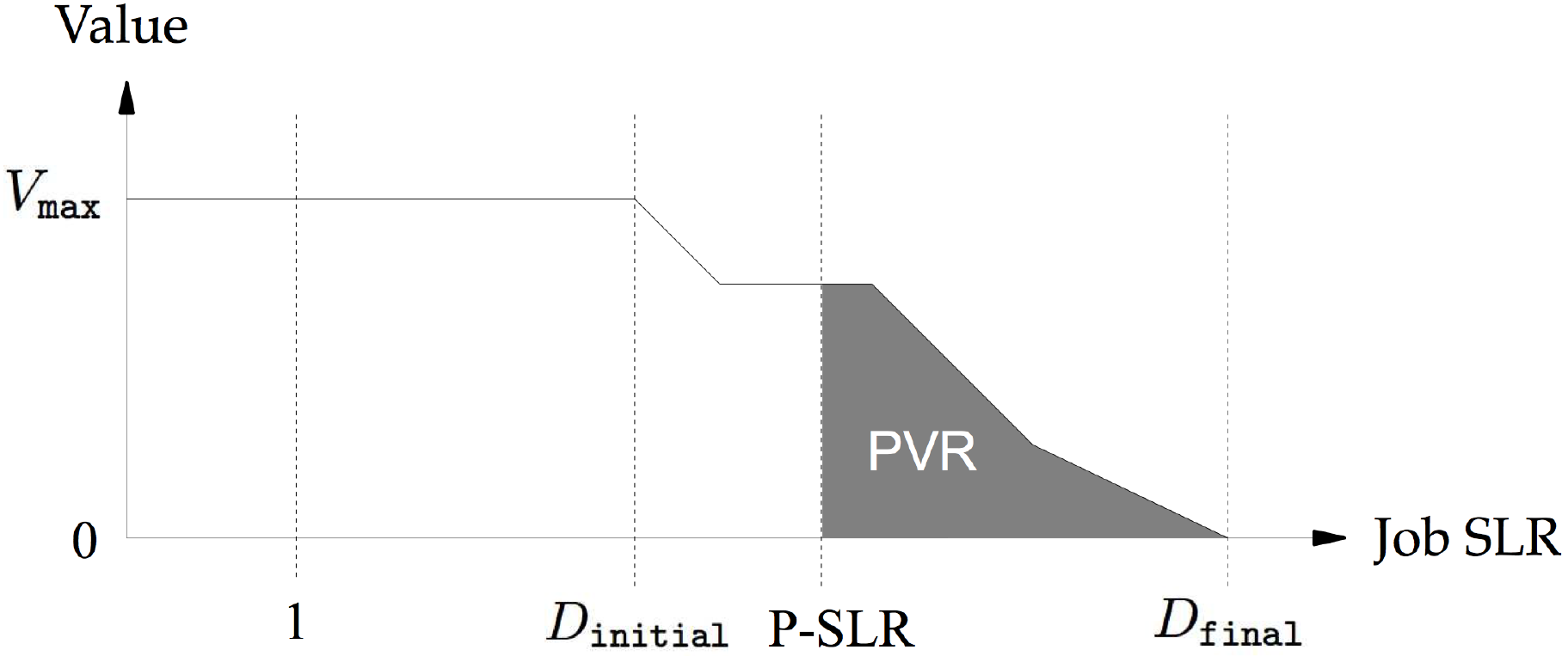}
\par\end{centering}

\protect\caption{Value Curve showing Projected Value Remaining }

\label{fig:pvr-diag}
\end{figure}

The Projected Value Remaining (PVR) policy \cite{aburkimsherEngD14}
is designed to capture the relative urgency of a task at a given moment.
The other policies may project tasks to be valuable or not, but are
unable to give an indication of whether that value is likely to decrease
with waiting much longer. PVR uses the P-SLR to determine the earliest
possible time a task could finish if it were run immediately. The
PVR is then the area under the value curve remaining between the P-SLR
at the current time and the final deadline $D_{\mathtt{final}}$ of
the job. This is illustrated graphically in Figure \ref{fig:pvr-diag}.

The tasks with the smallest value remaining are run first. Urgent
tasks would have a steeply sloping value curve, which would give only
a small area under the curve. Tasks about to time out would also have
only a small area remaining. Prioritising these tasks would reduce
starvation and loss of value.

The value curves were designed using linear interpolation between
the points so that the definite integral of this curve would be quickly
and exactly calculable using the trapezoidal method \cite{alvarado79}.
However, the policy method generalises to any value curves where value
can only decrease over time, as long as a final deadline is present.
The algorithm to calculate PVR is given in Algorithm \ref{projected-value-remaining-algo}.

\begin{algorithm}
$\mathtt{PVR}\left(T^{i},\:\mathtt{\tau_{current}}\right):$

\begin{eqnarray*}
J^{k} & = & T_{\mathtt{parent\_job}}^{i}\\
D_{\mathtt{final}} & = & J_{C_{D_{\mathtt{final}}}^{i}}^{k}\\
\mathtt{P\_SLR} & = & \frac{\left(T_{\mathtt{R}}^{i}+\tau_{\mathtt{current}}\right)-J_{\mathtt{arrive}}^{k}}{J_{\mathtt{CP}}^{k}}\\
PVR & = & \int_{\mathtt{P\_SLR}}^{D_{\mathtt{final}}}\mathtt{Value}\left(J^{k},s\right)\:ds\\
\mathtt{return} & PVR
\end{eqnarray*}

\protect\caption{Algorithm to compute Projected Value Remaining }

\label{projected-value-remaining-algo}
\end{algorithm}

\section{Experimental Method}

In order to be able to compare different schedules, appropriate metrics
are required. For this work, the pertinent metric is that of the proportion
of maximum value achieved. The maximum value of a workload is the
sum of the maximum value of all possible jobs, and is the value that
would be achievable if the number of processors available was unbounded,
there was no contention between work and no network delays. Because
different workloads may have different maximum values, it is necessary
to normalise these values between workloads. In this work, the value
achieved under a given set of circumstances is divided by the maximum
possible value to give a normalised figure.

Users are likely to not only care about their work, but also care
whether it was completed or not. Therefore the number of jobs starved
in a given set of circumstances will also be examined. A job is considered
to have starved (and is removed from the queue) once it has passed
its final deadline $D_{\mathtt{final}}$ as defined by its value curve.

The bidding policies are evaluated in simulation, using a synthetic
workload running over a synthetic platform. The simulations take place
within a discrete-time environment, implemented using the SimPy library
\cite{simpy} in the Python language. The platform consists of 4000
cores total, organised into clusters of 1000 cores each. Three clusters
are of architecture Kind1, and one of Kind2. The clusters are all
connected directly to the central router, where the auction is run.
Network delays between the clusters is accounted for by using a Communication
to Computation ratio of 0.2.

Ten synthetic workloads were used for evaluation. To ensure sufficient
scale was present, each workload was made up of 10,000 jobs with 5-20
tasks per job. This gives approximately 100,000 tasks in each workload.
The execution time distribution for jobs and tasks followed a log-uniform
distribution. 80\% of the tasks in the workload were Kind1, with 20\%
being of Kind2. The dependencies between jobs followed an exponential
distribution in node degree. These synthetic workloads were created
using the methods published by Burkimsher et al. \cite{burkimsher14}.
Value curves were also generated synthetically, and had $D_{\mathtt{initial}}$
values randomly selected between 2 and 4, $D_{\mathtt{final}}$ values
between 6 and 10, and 5-10 random, non-increasing points in between.

In simulation, real allocations of value are unavailable. Therefore,
it was assumed that the value of jobs was proportional to the number
of core-minutes their execution would consume. That is, that their
value density at $V_{\mathtt{max}}$ is identical. In periods of overload,
however, jobs may have very different value densities because the
value curves specified will lead to them having very different relative
urgency.

Load can be varied for the same workload by adjusting the inter-arrival
times of jobs, according to the method in Burkimsher et al. \cite{burkimsher14}.
The arrival rates are also adjusted to give peaks during working hours
and quieter periods overnight and at the weekend. This ensures that
many cycles of overload and catching-up are present, a more challenging
scenario for a scheduler than a constant arrival rate of work and
one that is more representative of real systems. Using this method,
load is varied between 70\% and 140\% of saturation, the state where
the system must operate at full capacity to service the work arriving.
Load above 100\% represents overload, where not all the work can be
run. With the daily and weekly cycles of load, transient overloads
will be present even below 100\% total load.

\section{Results and Discussion}

\begin{figure}
\begin{centering}
\includegraphics[clip,width=1\columnwidth]{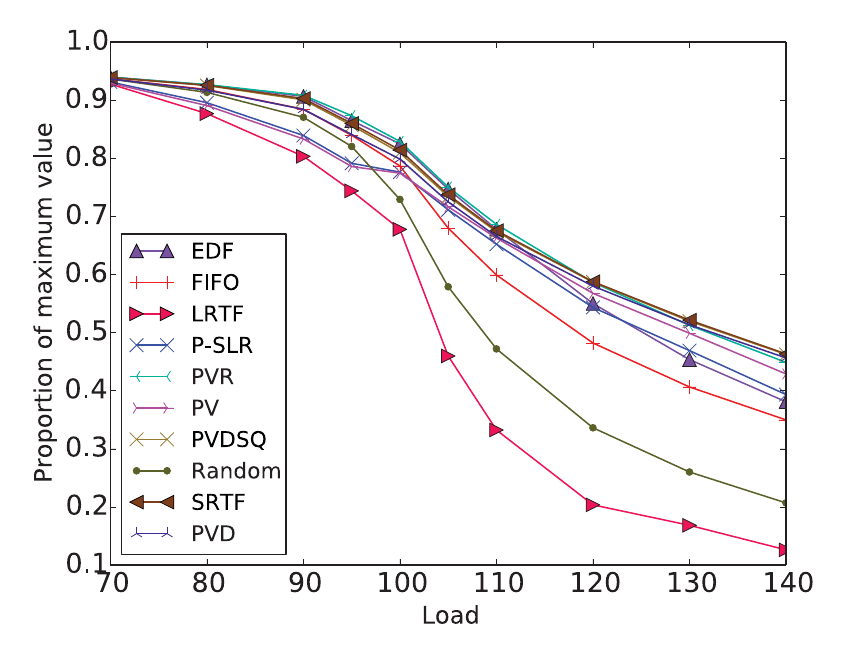}
\par\end{centering}

\protect\caption{Value against load for different bidding policies\label{fig:Value-against-load}}

\end{figure}

\begin{figure}
\begin{centering}
\includegraphics[clip,width=1\columnwidth]{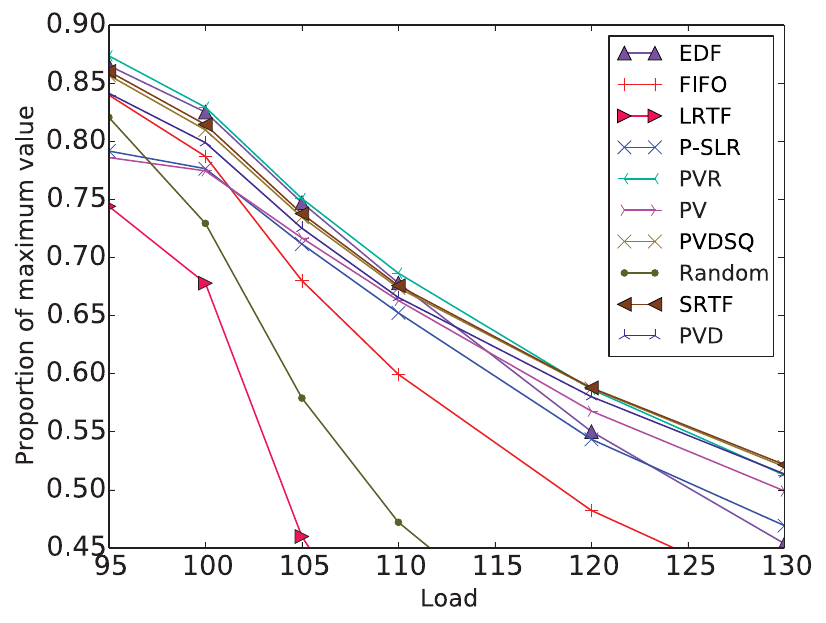}
\par\end{centering}

\protect\caption{Value against load for different bidding policies (Detail)\label{fig:Value-against-load-zoom}}
\end{figure}

\begin{figure}
\begin{centering}
\includegraphics[clip,width=1\columnwidth]{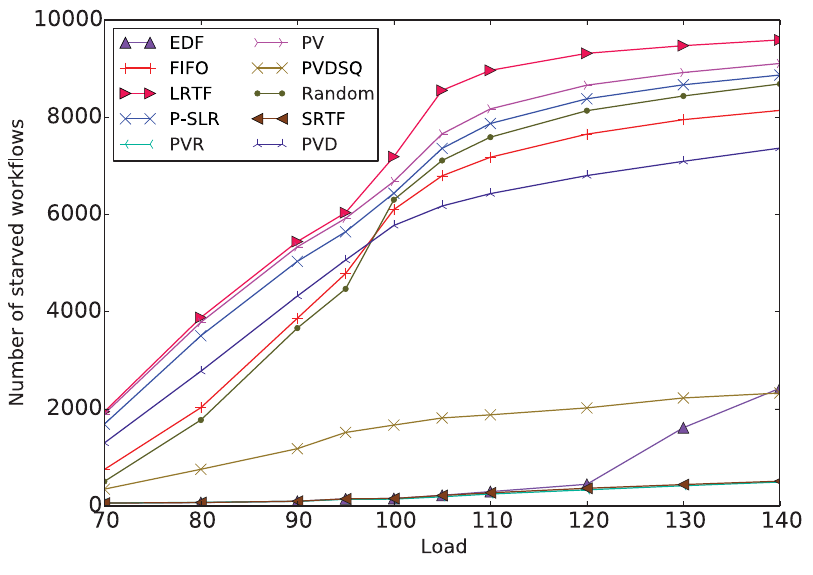}
\par\end{centering}

\protect\caption{Starvation against load for different bidding policies\label{fig:Starvation-against-load}}
\end{figure}

\begin{figure}
\begin{centering}
\includegraphics[clip,width=1\columnwidth]{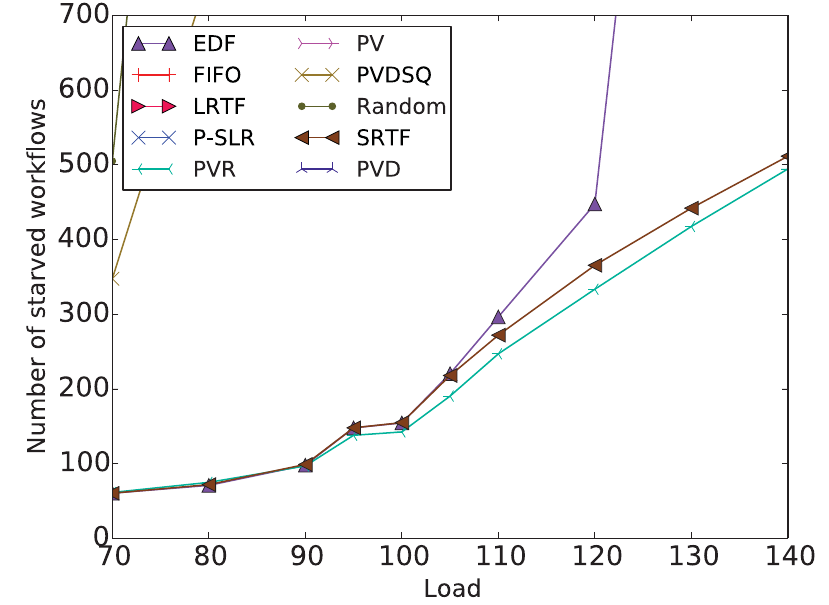}
\par\end{centering}

\protect\caption{Starvation against load for different bidding policies (Detail)\label{fig:Starvation-against-load-zoom}}
\end{figure}

\begin{figure}
\begin{centering}
\includegraphics[clip,width=1\columnwidth]{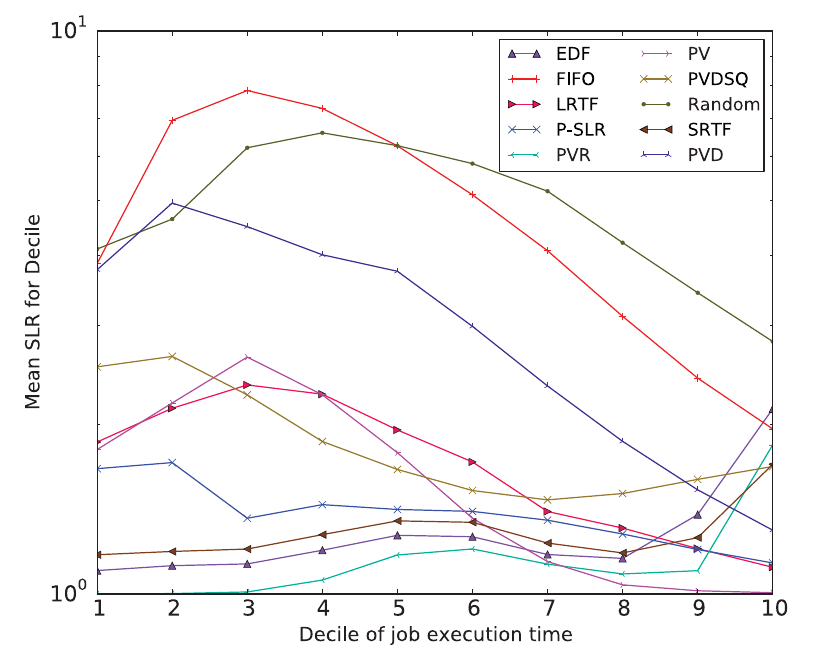}
\par\end{centering}

\protect\caption{Responsiveness (SLR) across the spectrum of job execution times at
120\% load\label{fig:Responsiveness-decile}}
\end{figure}

The experiments were set up such that there will always be transient
overloads, so that not all work is able to run. This is why the proportion
of maximum value is below 1.0 even at 70\% load. As load is increased,
the proportion of value decreases for all bidding policies. This is
bound to happen because the more work there is to do in a given time,
the more there will be that is not done. The challenge is to find
the policy that suffers the least loss of value given the increase
in load, with the results shown in Figure \ref{fig:Value-against-load}.

LRTF is a policy commonly used in static scheduling as it is good
at bin-packing, and hence at reducing workload makespan \cite{Topcuoglu2002}.
However, it is clear that it is a poor choice for a dynamic scheduling
scenario, because it will tend to run the most recently arrived jobs
first. This is because the jobs most recently arrived will have lots
of work left to do, and so be prioritised over those that are soon
to finish. In an overload situation, jobs will be started continuously
but never be able to finish. Its value achieved in this dynamic case
is even worse than the baseline random policy.

The FIFO policy is a fair improvement on the random policy, but FIFO
is still a long way behind the other policies under overload. This
is natural because it does not consider the size or the relative urgency
of jobs.

At load levels below overload, the PV policy performs surprisingly
poorly, even worse than the random policy. PV goes for the largest
and most valuable tasks first, meaning that many smaller ones will
starve. This affects the total value achievable, and it also leads
to the most tasks starving across the spectrum of load, with only
LRTF starving more. At very high levels of load, PV does not fall
as fast as some other policies, though it is still beaten by the flavours
of Value Density, SRTF and PVR.

P-SLR is designed to ensure that the SLR of jobs across the execution
time spectrum is equal (fairness with respect to responsiveness) \cite{burkimsher12}.
Under high levels of load, however, this can mean that many jobs may
have their final deadline before the SLR that is currently being achieved,
meaning that many will starve (Figure \ref{fig:Starvation-against-load}).
P-SLR is unable to take into account the fact that jobs may have varying
levels of urgency.

The two flavours of value density bidding (PVD and PVDSQ) are clearly
better than simply using value-based bidding. Locke \cite{Locke1986}
showed that for uniprocessor scheduling of workloads that are also
schedulable using EDF, value density based scheduling was optimal.
Naturally, workloads that overload their platform are not schedulable
using EDF, so these scenarios are outside those proved optimal. This
evaluation shows that in the overload scenarios considered, other
policies can do better than PVD.

Interestingly, PVDSQ gives higher value than PVD (Figure \ref{fig:Value-against-load}).
This is because PVDSQ starves many fewer tasks than PVD (Figure \ref{fig:Starvation-against-load}).
By squaring the value density, jobs that are not worth running are
more cleanly starved, leaving the rest to be able to run to completion.
PVDSQ also starves a few more of the largest jobs, which gives a lot
more space for other jobs to run. Interestingly, PVDSQ and SRTF give
almost indistinguishable levels of value across the load spectrum.
However, SRTF starves many fewer tasks overall because SRTF prioritises
small tasks with little time remaining.

EDF attains the second-highest level of value below 110\% load (Figure
\ref{fig:Value-against-load-zoom}). However, its performance falls
steeply after that. This is because once the earliest deadline that
is being run gets into the past, every job starts missing its deadline
and lateness starts cascading. Above 110\% load, the value attained
by EDF falls rapidly behind the alternative policies evaluated. After
120\% load, the number of tasks starved by EDF also increases rapidly
(Figure \ref{fig:Starvation-against-load-zoom}).

Below saturation and up to 110\% load, PVR attains the highest value
of all the policies evaluated. This is due to its ability to take
into account the urgency of tasks. Small, urgent tasks are able to
run, which ensures their value is captured before they starve. PVR
actually has the highest responsiveness (lowest SLR) of all the policies
evaluated for smaller jobs (Figure \ref{fig:Responsiveness-decile}).
It does this by only letting the largest jobs have their SLR increase.
However, the largest jobs can often also be those that are most tolerant
to waiting time: if a job takes months to run, it may be able to take
a few more days and users will not mind. This means that the value
curves are likely to decrease relatively gradually.

At very high levels of load, PVR's lead is eroded and it falls behind
SRTF and PVDSQ and is equalled by PVD at the highest level of load
evaluated. When there is too much work to run, PVR tends to keep most
jobs running by starving the largest ones. Yet the largest jobs can
also be the ones that supply the most value. As can be seen from Figure
\ref{fig:Responsiveness-decile}, SRTF penalises the largest jobs
a little less, meaning that it keeps more value overall under extreme
overload. Saying that, PVR has the lowest number of starved jobs across
the spectrum of load (Figure \ref{fig:Starvation-against-load-zoom}).
PVR manages to ensure that only 5\% of jobs are incomplete by their
final deadline, even at 140\% load. This low level of starvation is
likely to be attractive to grid operators because it will keep user
satisfaction high, as no user wishes to have their job starve.

\section{Conclusion}

In this paper, a number of bidding policies for market-based scheduling
in a distributed grid computing scenario were examined. The Projected
Value Remaining (PVR) policy was shown to achieve the highest levels
of value as long as overload was not excessive. Even under extreme
overload, the relatively low levels of starved jobs are likely to
be more acceptable to users than the slight improvement in overall
value offered by the PVDSQ policy. PVR is also likely to be preferable
to SRTF under extreme overload, because most jobs still have high
responsiveness, and it is the largest jobs, for which responsiveness
is usually less critical anyway because their execution times are
so long, that must wait longer. PVR's ability to capture urgency in
its bidding metric is what enables it to give these results.

A further notable conclusion is that the differences between the leading
policies are really quite close in terms of the value achieved, even
when the number of jobs they starve is wildly different. This evaluation
shows that different policies achieve similar levels of value using
quite different behaviour. Some policies run just the largest jobs
and starve everything else (PV), others the smallest (SRTF), while
others try to achieve balance across the space of execution times
(EDF, PVDSQ, PVR).

This work considers the scenario where all work is admitted but may
be left in the queue to starve. This may be less useful to users than
a system with an admission controller. Users could then know up-front
whether their jobs were accepted, and find out sooner if their jobs
were not going to be run. A natural direction for future work would
be to see whether integrating the current approaches with the addition
of an admission controller could achieve similar levels of value.

\smallskip{}

The research described in this paper is partially funded by the EU
FP7 DreamCloud Project (611411).

\bibliographystyle{plain}
\bibliography{biblob3}

\begin{thebibliography}{10}

\bibitem{aldarmi1999}
Saud~A. Aldarmi and Alan Burns.
\newblock Dynamic value-density for scheduling real-time systems.
\newblock In {\em Proceedings of The 11th Euromicro Conference on Real-Time
  Systems}, pages 270--277, June 1999.

\bibitem{alvarado79}
Fernando~L. Alvarado.
\newblock Parallel solution of transient problems by trapezoidal integration.
\newblock {\em IEEE Transactions on Power Apparatus and Systems},
  PAS-98(3):1080--1090, 1979.

\bibitem{burkimsher12}
Andrew Burkimsher, Iain Bate, and Leandro~Soares Indrusiak.
\newblock A survey of scheduling metrics and an improved ordering policy for
  list schedulers operating on workloads with dependencies and a wide variation
  in execution times.
\newblock {\em Future Generation Computer Systems}, 29(8):2009 -- 2025, October
  2013.

\bibitem{burkimsher14}
Andrew Burkimsher, Iain Bate, and Leandro~Soares Indrusiak.
\newblock A characterisation of the workload on an engineering design grid.
\newblock In {\em Proceedings of the High Performance Computing Symposium}, HPC
  2014, pages 8:1--8:8, San Diego, CA, USA, 2014. Society for Computer
  Simulation International.

\bibitem{aburkimsherEngD14}
Andrew~M. Burkimsher.
\newblock \textit{Fair, responsive scheduling of engineering workflows on
  computing grids}, {EngD Thesis}, {University of York}, 2014.

\bibitem{chen96}
Ken Chen and Paul Muhlethaler.
\newblock A scheduling algorithm for tasks described by time value function.
\newblock {\em Real-Time Systems}, 10(3):293--312, 1996.

\bibitem{dube08}
Nicolas Dube and Marc Parizeau.
\newblock Utility computing and market-based scheduling: Shortcomings for grid
  resources sharing and the next steps.
\newblock In {\em Proceedings of the 22nd International Symposium on High
  Performance Computing Systems and Applications, 2008}, HPCS 2008, pages
  59--68, June 2008.

\bibitem{graham69}
Ronald~L. Graham.
\newblock Bounds on multiprocessing timing anomalies.
\newblock {\em SIAM Journal on Applied Mathematics}, 17:416--429, 1969.

\bibitem{intel04}
{Intel Corporation}.
\newblock Enhanced {Intel} {SpeedStep} technology for the {Intel} {Pentium M}
  processor - white paper, March 2004.

\bibitem{irwin04}
David~E. Irwin, Laura~E. Grit, and Jeffrey~S. Chase.
\newblock Balancing risk and reward in a market-based task service.
\newblock In {\em Proceedings of the 13th IEEE International Symposium on High
  Performance Distributed Computing}, HPDC 2004, pages 160--169, Washington,
  DC, USA, 2004. IEEE Computer Society.

\bibitem{kelley61}
James~E. {Kelley, Jr.}
\newblock Critical-path planning and scheduling: Mathematical basis.
\newblock {\em Operations Research}, 9(3):pp. 296--320, 1961.

\bibitem{kesselman98}
Carl Kesselman and Ian Foster.
\newblock {\em The Grid: Blueprint for a New Computing Infrastructure}.
\newblock {Morgan Kaufmann Publishers}, San Francisco, CA, USA, November 1999.

\bibitem{Lai05admissons}
Kevin Lai, Lars Rasmusson, Eytan Adar, Li~Zhang, and Bernardo~A. Huberman.
\newblock Tycoon: An implementation of a distributed, market-based resource
  allocation system.
\newblock {\em Multiagent and Grid Systems}, 1(3):169--182, August 2005.

\bibitem{Lee99}
Chen Lee, John Lehoczky, Dan Siewiorek, Ragunathan Rajkumar, and Jeff Hansen.
\newblock A scalable solution to the multi-resource {QoS} problem.
\newblock In {\em Proceedings of the 20th IEEE Real-Time Systems Symposium
  (RTSS 1999)}, pages 315--326, Washington, DC, USA, 1999. IEEE Computer
  Society.

\bibitem{Locke1986}
Carey~Douglass Locke.
\newblock {\em Best-effort decision-making for real-time scheduling}.
\newblock PhD thesis, Carnegie-Mellon University, Pittsburgh, PA, USA, May
  1986.

\bibitem{Maheswaran99}
Muthucumaru Maheswaran, Tracy~D. Braun, and Howard~Jay Siegel.
\newblock Heterogeneous distributed computing.
\newblock In {\em Encyclopedia of Electrical and Electronics Engineering},
  pages 679--690. John Wiley, 1999.

\bibitem{millerdrexler88}
Mark~S. Miller and K.~Eric Drexler.
\newblock Markets and computation: Agoric open systems.
\newblock In B.~A. Huberman, editor, {\em The Ecology of Computation}, pages
  133--176, Amsterdam, 1988. North-Holland Publishing Company.

\bibitem{popoviciwilkes05}
Florentina~I. Popovici and John Wilkes.
\newblock Profitable services in an uncertain world.
\newblock In {\em Proceedings of the ACM/IEEE SC 2005 Conference on
  Supercomputing}, pages 36--36, November 2005.

\bibitem{saule10}
Erik Saule, Doruk Bozda\u{g}, and Umit~V. Catalyurek.
\newblock A moldable online scheduling algorithm and its application to
  parallel short sequence mapping.
\newblock In Eitan Frachtenberg and Uwe Schwiegelshohn, editors, {\em Job
  Scheduling Strategies for Parallel Processing}, volume 6253 of {\em Lecture
  Notes in Computer Science}, pages 93--109. Springer Berlin Heidelberg, 2010.

\bibitem{simpy}
Team SimPy.
\newblock Welcome to simpy, 2015.

\bibitem{Sutherland68}
Ivan~Edward Sutherland.
\newblock A futures market in computer time.
\newblock {\em Communications of the ACM}, 11(6):449--451, June 1968.

\bibitem{Topcuoglu2002}
Haluk Topcuouglu, Salim Hariri, and Min-You Wu.
\newblock Performance-effective and low-complexity task scheduling for
  heterogeneous computing.
\newblock {\em IEEE Transactions on Parallel and Distributed Systems},
  13(3):260--274, March 2002.

\bibitem{cloudintro11}
William Voorsluys, James Broberg, and Rajkumar Buyya.
\newblock {\em Cloud Computing: Principles and Paradigms}, chapter Introduction
  to Cloud Computing, pages 1--44.
\newblock Wiley Press, New York, USA, February 2011.

\bibitem{waldspurger02}
Carl~A. Waldspurger, Tad Hogg, Bernardo~A. Huberman, Jeffrey~O. Kephart, and
  W.~Scott Stornetta.
\newblock Spawn: a distributed computational economy.
\newblock {\em IEEE Transactions on Software Engineering}, 18(2):103--117,
  February 1992.

\bibitem{zhaosakellariou06}
Henan Zhao and Rizos Sakellariou.
\newblock Scheduling multiple {DAGs} onto heterogeneous systems.
\newblock In {\em Proceedings of the 20th International Parallel and
  Distributed Processing Symposium (IPDPS)}, 2006.

\end{thebibliography}

\end{document}